\documentclass[twocolumn]{aastex631}

\usepackage{chngpage}
\usepackage{booktabs}
\usepackage{amsmath}
\usepackage{xcolor}

\newcommand\RGMX{\bgroup\markoverwith{\textcolor{cyan}{\rule[0.5ex]{4pt}{1pt}}}\ULon}

\newcommand\ACCX{\bgroup\markoverwith{\textcolor{red}{\rule[0.5ex]{4pt}{1pt}}}\ULon}
\shorttitle{Life around M-dwarfs}
\shortauthors{Childs, Martin \& Livio}
\graphicspath{{./}{figures/}}

\begin{document}

\title{Life on Exoplanets In the Habitable Zone of M-Dwarfs?}

\author[0000-0002-9343-8612]{Anna C. Childs}
\author[0000-0003-2401-7168]{Rebecca G. Martin}
\author{Mario Livio}
\affiliation{Nevada Center for Astrophysics, University of Nevada, Las Vegas, NV 89154, USA}
\affiliation{Department of Physics and Astronomy, University of Nevada, Las Vegas, 4505 South Maryland Parkway,
Las Vegas, NV 89154, USA}



\begin{abstract}
Exoplanets orbiting in the habitable zone around M-dwarf stars have been prime targets in the search for life due to the long lifetimes of the host star, the prominence of such stars in the galaxy, and the apparent excess of terrestrial planets found around M-dwarfs.  However, the heightened stellar activity of M-dwarfs and the often tidally locked planets in these systems have raised questions about the habitability of these planets. In this letter we examine another significant challenge that may exist: these systems seem to lack the architecture necessary to deliver asteroids to the habitable terrestrial planets, and asteroid impacts may play a crucial role in the origin of life.  The most widely accepted mechanism for producing a stable asteroid belt and the late stage delivery of asteroids after gas disk dissipation requires a giant planet exterior to the snow line radius.  We show that none of the observed systems with planets in the habitable zone of their star also contain a giant planet and therefore are unlikely to have stable asteroid belts.  We consider the locations of observed giant planets relative to the snow line radius as a function of stellar mass and find that there is a population of giant planets outside of the snow line radius around M-dwarfs. Therefore, asteroid belt formation around M-dwarfs is generally possible. However, we find that multi-planetary system architectures around M-dwarfs can be quite different from those around more massive stars. 

\end{abstract}


\section{Introduction} \label{sec:intro}
The habitable-zone (HZ) is commonly defined as the range of circumstellar distances from a star within which a rocky planet may have liquid water on its surface, given a dense enough atmosphere \citep{Kasting1993}.  Many of the HZ Earth-size exoplanets discovered by the Kepler and TESS missions were found to orbit M-dwarf stars, and a
number of those have become obvious targets for at least partial
atmospheric characterization by JWST \citep{Seager2010, Batalha2013, Dressing2015, Anglada-Escud2016, Gillon2017}.  This brought into focus again the important
question of whether planets orbiting M-dwarfs can harbor life \citep[e.g.][]{Shields2016, Wandel2018}.

The point is that in spite of their seemingly positive attributes vis-à-vis the question of life, several concerns have been raised over the
suitability of M-dwarfs as hosts to life-bearing planets. Here are just a
few of those. First, M-dwarfs often experience intense flaring activity which sometimes also trigger mass ejections \citep{Hawley1991, Lammer2013}. The flares increase in both
frequency of occurrence and amplitude the smaller the mass of the star \citep{Hilton_2010, Davenport_2012}.  Even M-dwarfs which are relatively less active exhibit flares and
episodes of significant UV and X-ray emission during the first billion years or so of their evolution \citep{France_2013, France_2016}. Planets exposed to these harsh events
could lose their atmospheres, their oceans, or both, by the time the star has settled onto its long main-sequence phase, and be totally sterilized \citep{Luger2015, Bomont2017,Tilley2019}.
Second, since the habitable zone around M-dwarfs is much closer in to the central star, planets are likely to be tidally locked \citep{Kasting1993, Barnes2017}. This could generate a huge temperature difference between the permanent day and perpetual night sides of the exoplanet, with gases on the night side being
frozen solid, while the day side is being star-baked dry. As a result, there
used to be considerable skepticism on whether tidally locked planets could host a biosphere, since the impression was that it would have had to be confined to that eternal, narrow, twilight zone around the
terminator separating the two sides \citep{Kite_2011, Wordsworth_2011, Checlair_2019}.

Opinions on the habitability of planets orbiting M-dwarfs started to
change in recent years. First, it has been suggested that magnetic fields could shield exoplanets from the deleterious effects of flares \citep{Segura2010, Kay2016}, and prevent a significant erosion of the atmospheres and oceans by mitigating against the effects of stellar winds \citep{Ward2000}. Second, theoretical
simulations identified atmospheric mechanisms that could, in the case of
sufficiently dense atmospheres, circulate and distribute heat from the day
side to the night side \citep[e.g.][]{Sergeev_2020}. In
addition, modeling has shown that airborne mineral dust could cool the
day side and warm the night side of a tidally locked planet, thereby
broadening the habitable area \citep{Boutle_2020}. A feedback
mechanism that could increase the amount of dust in the atmosphere and delay the loss of water from the day side on planets at the inner edge of the HZ has also been identified \citep{Boutle_2020}. As a result of these and similar
theoretical ideas, this class of objects has become the main target in the search for extrasolar life, even though some questions remain (e.g.
regarding the bioactive UV fluence; \cite{Ranjan_2017}).  

While these and other studies are optimistic about the possibility of harboring life on habitable exoplanets around M-dwarfs, most works have ignored the potential role that asteroid
impacts on the early Earth may have played in originating life; although
see an extensive review by \cite{Osinski2020} and study by \cite{Grazier2016}. In particular, recent
work suggests that large impacts during the “late veneer” (after the formation of the Moon) may have been crucial for producing a reduced atmosphere on the early Earth \citep{Sinclair2020}. This would have resulted from the
reaction of the iron core of the impactor with water in the oceans — as
the iron oxidized, hydrogen was released, producing an atmosphere
favorable for the emergence of simple organic molecules \citep[e.g.][]{Zahnle2019, Genda_2017, Benner_2020} (and see \citet{Sutherland2017} and \citet{Szostak2017} for current thoughts on origin of life).

In the present work, we therefore examine (in Section \ref{sec:HZplanets}) whether
the conditions that may be necessary to bring about similar impacts onto
terrestrial exoplanets are satisfied for all known HZ exoplanets orbiting M-dwarfs. We also investigate the occurrence rates of giant planets around low-mass stars and whether asteroid impacts are to be expected onto exoplanets hosted by M-dwarfs in general in Section \ref{sec:architectures}. A
discussion and conclusions follow.

\section{Planets in the habitable zone}
 \label{sec:HZplanets}
\begin{figure*}
\centering
		\includegraphics[width=1\columnwidth]{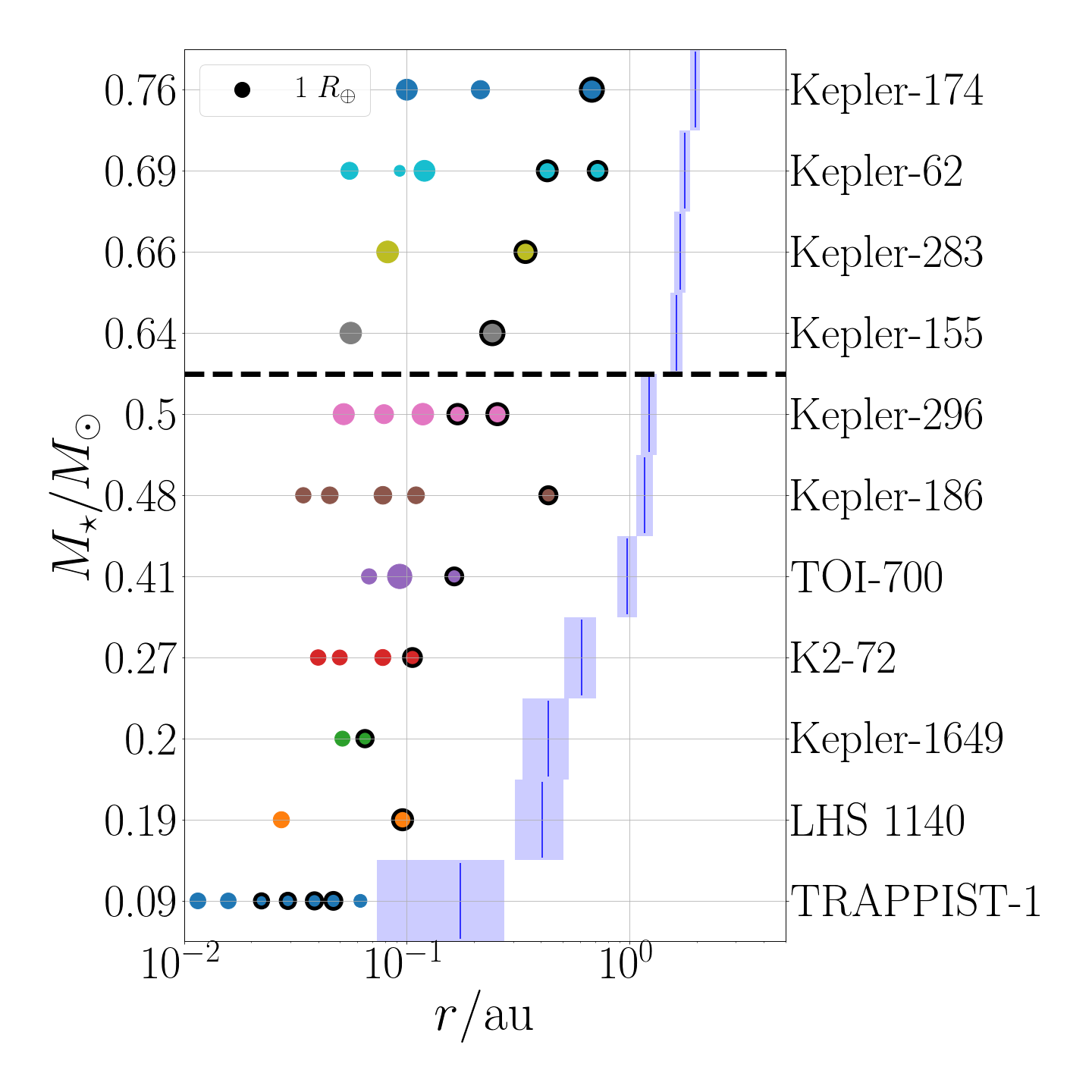}
		\includegraphics[width=1\columnwidth]{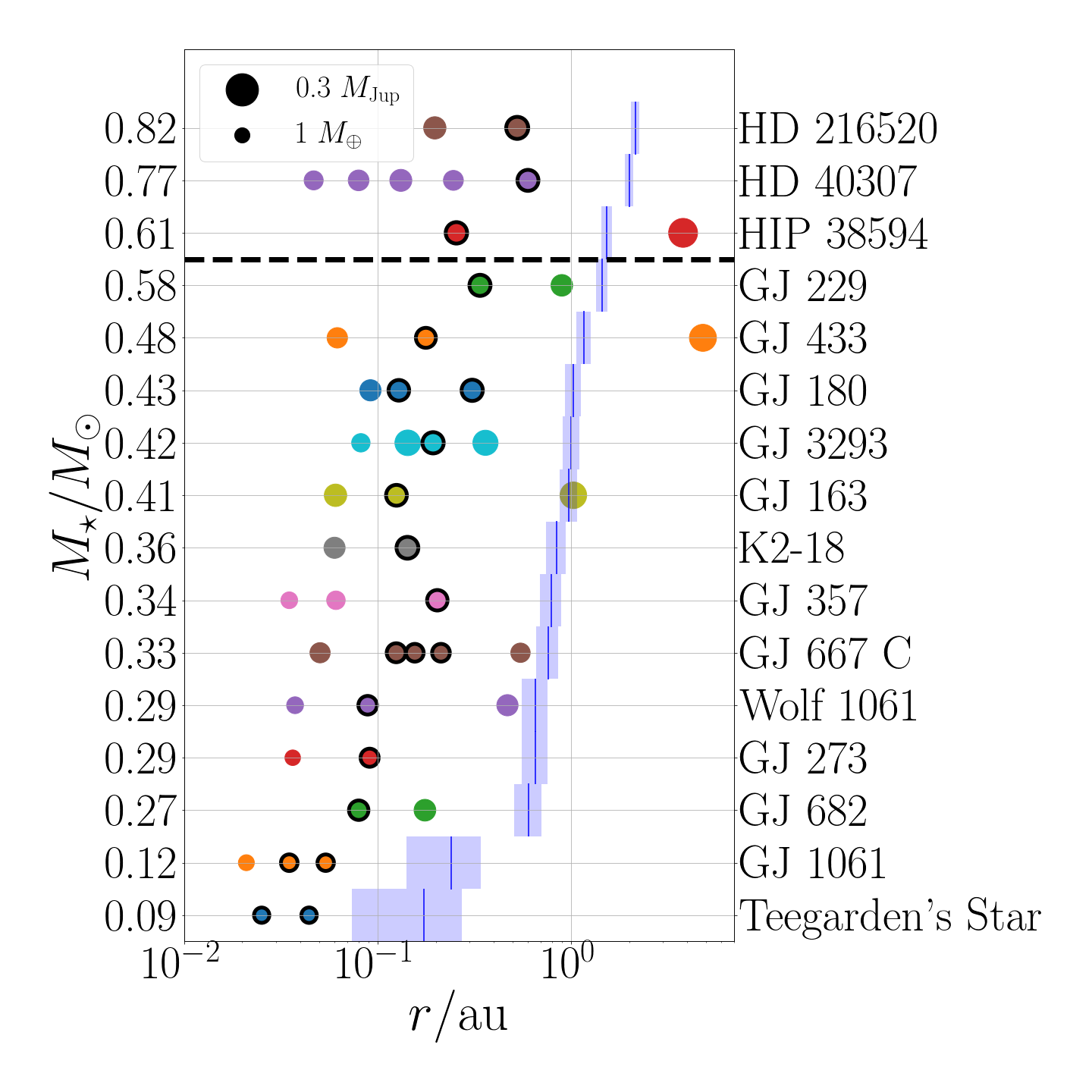}
    \caption{Observed multi-planetary systems that harbor a terrestrial planet in the HZ of their host star.  Systems with radius measurements are shown on the left and systems with mass measurements are shown on the right.  In each plot, the system name is listed on the right and the stellar mass on the left.  The size of each point is scaled to the planet radius (left) or to $m^{1/3}$ (right), where $m$ is the planet mass.  The size of an Earth-size planet is shown in the top left corner for reference. The location of the snow line radius is marked by a vertical blue line and the $1\sigma$ bounds are shaded in light blue.  The planets in the habitable zone of the system are highlighted with a black ring.  M-dwarfs are separated from the K and G-type stars by a black dashed line.}
    \label{fig:HZ_planets}
\end{figure*}
Currently, there are 48 distinct systems with confirmed terrestrial-like planets--planets with a mass less than $10 \, M_{\oplus}$, or are smaller than $2.5 \, R_{\oplus}$, orbiting M, K, or G-type stars in the systems HZ.  The habitability estimates are taken from the Habitable Exoplanets Catalog (HEC) \footnote{UPR Habitable Exoplanets Catalog: \url{https://phl.upr.edu/projects/habitable-exoplanets-catalog} queried on August 29, 2022.}.  Of these 48 systems, 27 systems contain more than one planet.  Figure \ref{fig:HZ_planets} shows these 27 systems, using data from the NASA Exoplanet Archive \footnote{NASA Exoplanet Archive: \url{https://exoplanetarchive.ipac.caltech.edu} queried on August 29, 2022 using the TAP Interface to Planetary Systems Data tool.  We report default planetary values.} \citep{NASA_Exoplanet_Science_Institute2020-gj}.  11 of these multi-planet systems contain radius measurements for all of the system planets and 16 have mass measurements for all of the system planets.  Multi-planet systems with radius measurements are shown on the left and multi-planet systems with mass measurements are shown on the right.  If only the orbital period is provided for a planet, we use Kepler's Third Law
to find the semi-major axis, $r$.
M-dwarfs (stars with $M_{\star}\le 0.6 \, M_{\odot}$) are separated from the K and G-type stars by a black dashed line.  
\begin{figure*}
\centering
	\includegraphics[width=2\columnwidth]{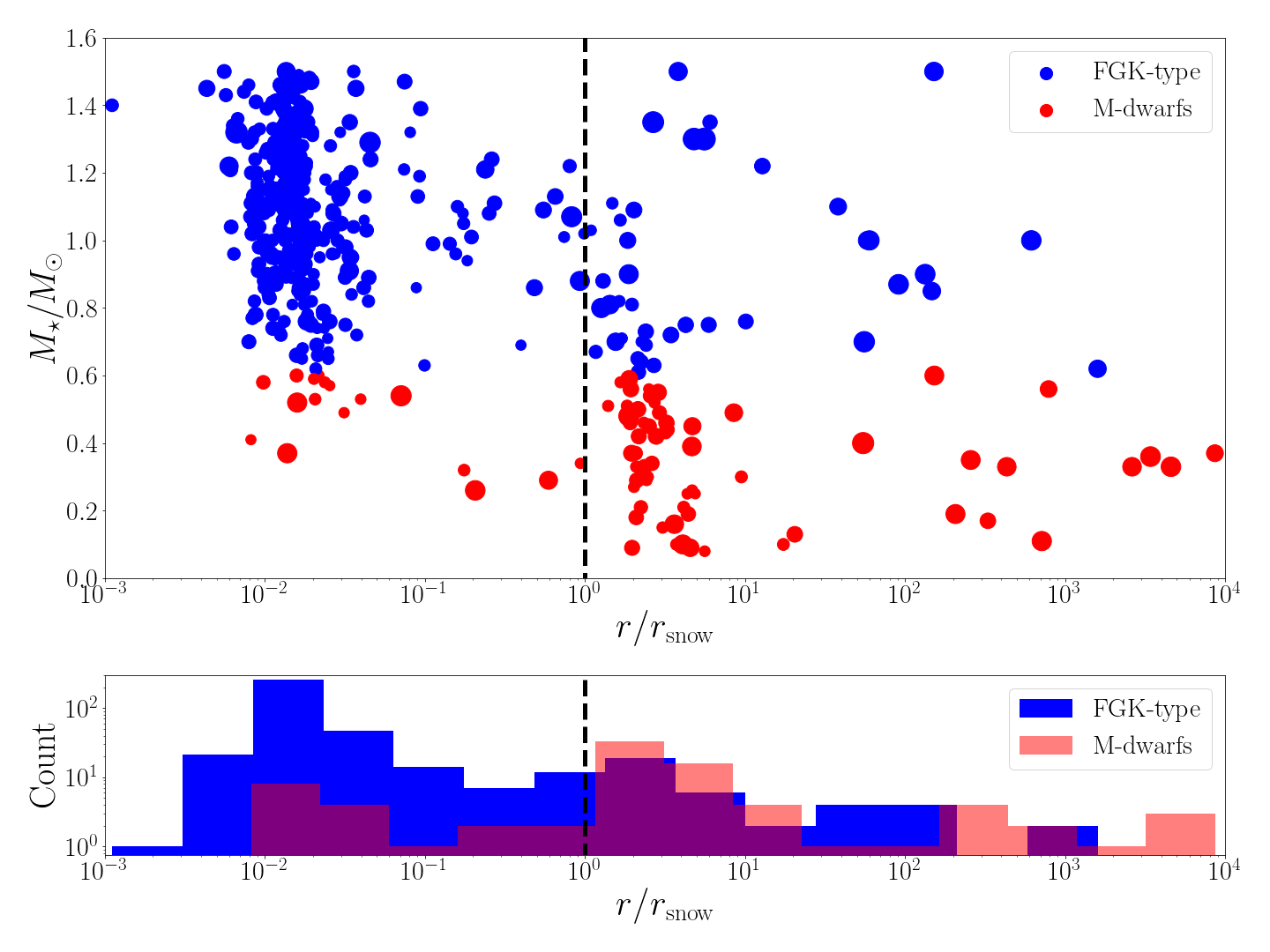}
    \caption{Locations of the giant planets, $r$, normalized by the snow line radius in the system, versus the stellar mass, $M_\star$.  The point sizes in the top plot are proportional to $m^{1/3}$.  Red dots indicate planets around M-dwarf stars and blue dots indicate planets around FGK-type stars.  The point sizes in the legend correspond to Jupiter mass planets.  The bottom plot shows normalized histograms of the giant planet locations for both single planet and multi-planet systems.  The location of the snow line is marked by a black dashed vertical line.}
    \label{fig:giant_locations}
\end{figure*}
To estimate the location of the snow line as a function of stellar mass we use 
\begin{equation}\label{eq:Sun_snow}
    r_{\mathrm{snow}} \approx 2.7 \left ( \frac{M}{M_{\odot}} \right)^{1.14} \, \rm au
\end{equation}
\citep{Mulders2015b}.  The coefficient of 2.7, which comes from normalizing to the solar system, is adopted from \cite{Martin2013}.  The location of the snow line radius is marked by a vertical blue line and the $1\sigma$ bounds, $\pm 0.1 \, \rm au$, are shaded in light blue.  The snow line marks the radial distance at which water freezes \citep{Hayashi1981}.  The freezing of water results in an increase in solid material at and exterior to the snow line.  Consequently, stable asteroid belts are generally expected to form near the snow line, interior to giant planets that form exterior to it \citep{Martin2013,Ballering2017} and giant planets are expected to form exterior to the snow line \citep{Pollack1996, Kennedy2008}.

We adopt the definition of a giant planet from \cite{Hatzes_2015} as a planet having a mass of $0.3-60 \, M_{\rm Jup}$.  There are no known giant planets in any of the systems with planets in the HZ \footnote{The HEC has listed GJ 832 c as a habitable planet that does have a gas giant companion exterior to the snowline.  However, \cite{Gorrini2022} have recently reassessed the RV measurements using updated stellar parameters.  Their results negate the existence of the habitable planet and so we do not include this system in our Figure \ref{fig:HZ_planets}.}.  Observational biases against planets on long period orbits render observations incomplete and so it is possible the data are missing exterior giant planets.  However, we are interested in giant planets that are not too far away from the snow line around M-dwarfs.  The snow line radius increases with stellar mass, but around M-dwarfs it is located in the approximate range $0.2-1.2\,\rm au$.  RV surveys are more likely to detect giant planets around low mass stars than around higher mass stars since a low mass star will move more which allows giant planets on longer orbits to be detected.

The Kepler detection limit is at orbital periods near 200 days due to the criterion that three transits need to be observed in order for a planet to be confirmed  \citep{Bryson_2020}. However, in the case of low signal-to-noise observations, two observed transits may suffice which allows longer period orbits to be detected.  This was the case for Kepler-421 b which has an orbital period of 704 days \citep{Kipping2014}.  Furthermore, any undetected exterior giant planets would likely raise a detectable transit timing variation (TTV) signal on the inner planets \citep{Agol2004}.  For these reasons, while the observations could be missing long period giant planets, the lack of giant planets around low mass stars that are not too far from the snow line is likely real.  Because the systems that contain a planet in the HZ zone likely do not contain a giant planet not too far from the snow line, we suggest that these systems do not contain stable asteroid belts.
The most widely accepted mechanism for producing stable asteroid belts requires perturbations from external giant planets which inhibit the formation of a planet in the belt region \citep[e.g.][]{Petit2001,Martin2013}.  

The $\nu_6$ secular resonance has been demonstrated to be the main channel for the late-stage delivery of asteroids to Earth, after gas disk dissipation \citep{Morbidelli2000,Ito2006,Smallwood_2017,Martin2022asteroids}.  The $\nu_6$ secular resonance is the most prominent resonance in the solar system and it is the result of interactions between the apsidal precession rate of Saturn and the asteroid belt \citep{Froeschle1986,Morbidelli1994,Ito2006, Minton_2011}. In order for a planetary system to have such a secular resonance that can deliver material from the asteroid belt to a HZ planet, requires two giant planets outside of the snow line radius.  If there are no giant planets, there are no resonances and so there is no instability in a belt, even if it can form.  With $n$-body simulations of an asteroid belt comprised of test particles, \cite{Martin2022asteroids} demonstrated that if Saturn and Jupiter are in a 2:1 mean motion resonance, while the asteroid belt is unstable, there is  a significant reduction of asteroid delivery to Earth because there is no $\nu_6$ resonance.  Other mechanisms that may result in asteroid collisions with inner terrestrial planets include asteroid-asteroid interactions and the Yarkovsky Effect. However, these mechanisms operate on longer timescales and may be unlikely to result in a sufficient delivery of asteroids to the HZ planets \citep{Greenberg2017}. 

Whilst there are no giant planets, the most massive planet found in the HZ systems in Figure \ref{fig:HZ_planets} is the Super-Neptune HIP 38594 c with a mass of $0.152 \pm 0.023 \, M_{\rm Jup}$ \citep{Feng2020}.  Three systems, HIP 38594, GJ 433, and GJ 163, contain a Super-Neptune exterior to the snowline.  \cite{Childs2019} modeled terrestrial planet formation in the solar system using different masses for Jupiter and Saturn.  They found that $15 \, M_{\oplus}$ and $45 \, M_{\oplus}$ planets at Saturn's and Jupiter's orbit respectively, take more than ten times longer than Jupiter and Saturn to eject $10 \%$ of the initial disk material from the system.  These findings indicate that small Super-Neptunes may not be massive enough to sustain an asteroid bombardment similar to what the early Earth experienced in the solar system.

The lack of giant planets in the (so far) observed systems containing HZ exoplanets suggests that these systems are unlikely to harbor an asteroid belt and the mechanism required for late-stage asteroid delivery to the HZ.  Therefore, \textit{if} asteroid impacts are indeed necessary for life, it is unlikely that the observed planets in the HZ harbor life.

\section{Observed planetary system architectures}\label{sec:architectures}
Exoplanet observations suggest that small planet occurrence rates decrease with increasing stellar mass \citep[e.g.][]{Mulders2015, Hardegree-Ullman2019} and giant planet occurrence rates increase with stellar mass \citep[e.g.][]{Endl2006, Johnson2007,Johnson2010, Bonfils2013, Obermeier2016, Ghezzi2018}. Furthermore, an excess of super-Earths has been observed around M-dwarfs.  It is thought that these super-Earths form via pebble-accretion.  The excess of these super-Earths may indicate the absence of giant planets as giant planets exterior to the snow line would decrease the pebble flux to the inner disk regions, inhibiting the growth of an inner super-Earth \citep{Mulders_2021}.  Similarly, models of pebble accretion are able to reproduce systems that are similar to exoplanet observations, but it is unable to produce systems that resemble the solar system \citep{Izidoro2021}.

The majority of transiting planets have been observed by Kepler, whose main targets were Sun-like stars.  The Kepler targets were magnitude limited and as a result only about $2\%$ of target stars were M-dwarfs \citep{Gaidos2016}.  Additionally, both radial velocity (RV) and transit techniques are not sensitive to longer period planets, specifically the giant planets beyond their systems snow line, and both techniques struggle with the intrinsic low luminosity of M-dwarfs.  

 While more giant planets are found around more massive stars, giant planets are still found around M-dwarfs.  \cite{Schlecker2022} recently reassessed planet occurrence rates around M-dwarfs after accounting for observational biases associated with RV techniques.   They simulated synthetic data using a core-accretion with migration model and compared the synthetic data to observational RV data from the High Accuracy RV Planet Searcher (HARPS, \cite{Mayor2003}) and the Calar Alto high-Resolution search for M dwarfs with Exoearths with Near-infrared and optical Echelle Spectographs (CARMENES, \cite{Quirrenbach2010, Reiners2018}).  They found that the simulated data are consistent with observational data for small planets around low-mass stars.  However, the observations reveal an excess of giant planets around low-mass stars that is not reproduced in the synthetic population.  Still, the trend of increasing giant planet occurrence rate with stellar mass holds. 
 
 The excess of giant planets that cannot be explained by core-accretion models may imply a separate formation pathway for giant planets around low-mass stars.  The question therefore rises: can the formation pathway of giant planets around low-mass stars produce systems with giant planets exterior to the snow line, an asteroid belt, and an inner terrestrial system?

\subsection{Giant planet locations relative to the snow line}

To begin to answer this question we first consider how the locations of observed giant planets compare to the snow line radius for different stellar masses.  The top panel of Figure \ref{fig:giant_locations} shows the semi-major axis, $r$, of all giant planets normalized by the snow line radius in the system versus the stellar mass.   We now consider systems around stars in the mass range $0.08-1.5 \, M_{\odot}$, which includes M,K,G and F-type stars. In this stellar mass range, life may have sufficient time to evolve on habitable planets.  Planets around M-dwarfs are marked in red and planets around K,G, and F-type stars are marked in blue.  The bottom panel shows the histograms for the giant planet locations in FGK-type systems (blue) and M-dwarf (red) systems.  We mark the location of the snow line by a black dashed vertical line.

A clear trend emerges: giant planets around M-dwarfs are largely found exterior to the snow line.
We find that there is a peak in the giant planet occurrence rate around the snow line for both low and high mass stars. This is consistent with the findings of \cite{Fernandes_2019} who considered solar-mass stars.
While there exists an observational bias against the long period giant planets that may exist near the snow line around G and F-type stars, the peak around the snow line radius and the lack of hot and warm Jupiters around M-dwarfs is likely real (see \cite{Obermeier_2016} for a more in-depth discussion on the apparent lack of hot Jupiters around M-dwarfs).  This trend may point to potential differences in giant planet formation and/or evolution between those around low-mass and higher mass stars (see the following section for a discussion on formation theories).


The population of giant planets outside of the snow line radius around M-dwarfs suggests that asteroid belt formation around M-dwarfs is generally possible.  
 Warm dust belts may be indicative of asteroid belts in terrestrial planet regions \citep{Martin2013,Ballering2017} or evidence of giant impacts  \citep[e.g.][]{Morales2011,Moor2021}. \cite{Theissen2014} found evidence for warm dust around a sample of M-dwarfs (see their table 5 for the minimum orbital distance of the dust). Currently, there is no evidence of even a Kuiper belt equivalent in TRAPPIST-1 \citep{Marino2020} although in general, cold debris disks are found around a similar fraction of M-dwarfs as higher mass stars \citep{Lestrade2006,Luppe2020}.

\subsection{Multi-planet systems with a giant planet}

\begin{figure*}
\centering
	\includegraphics[width=2\columnwidth]{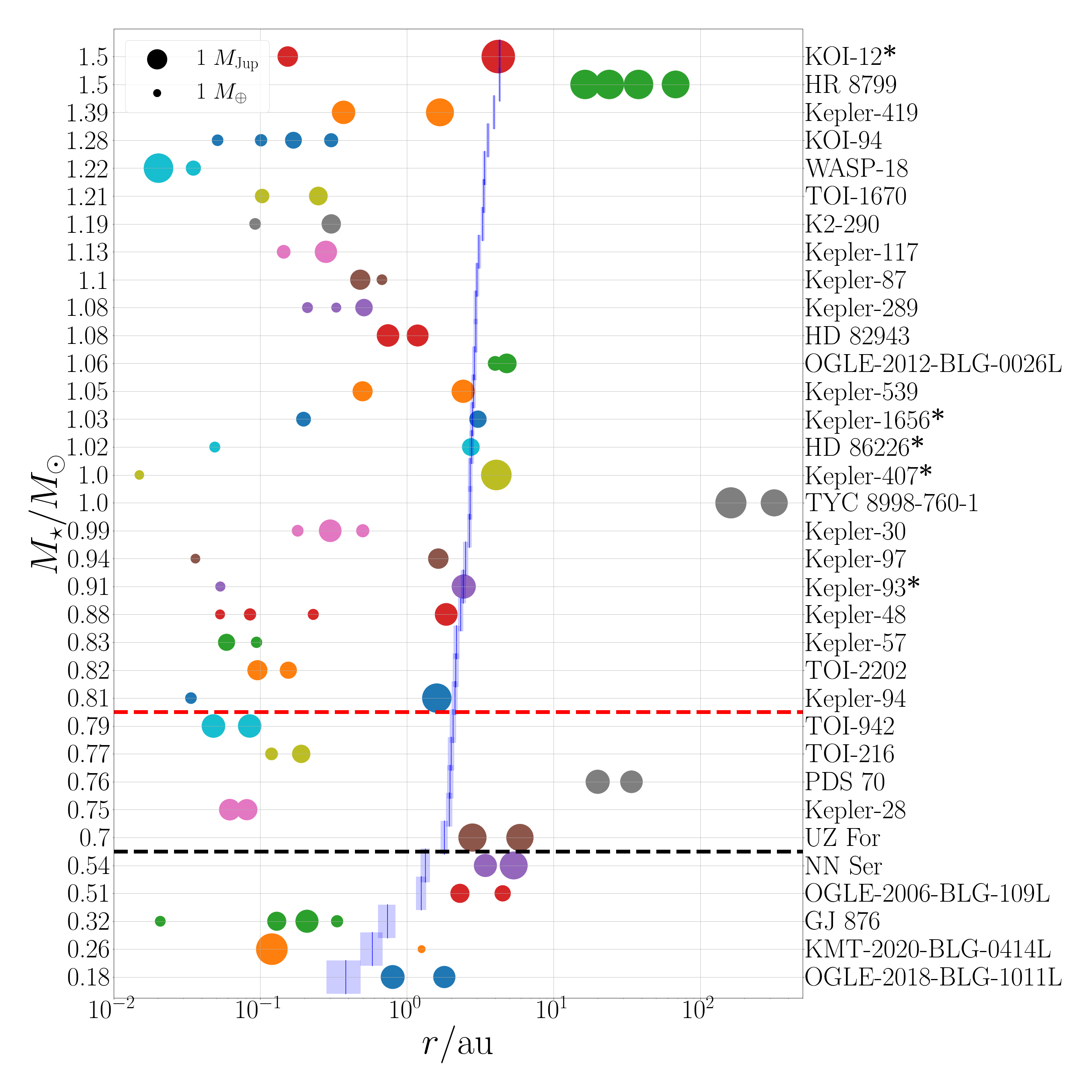}

    \caption{Multi-planet systems with mass measurements that contain a giant planet. The system name is listed on the right and the stellar mass on the left. The point sizes are proportional to $m^{1/3}$.  We show the size of Jupiter and Earth mass planets in the top left corner for reference.  We plot the location of the snow line with a blue vertical line and shade the $1\sigma$ bounds of the snow line radius in light blue.  The red dashed line separates the M-dwarfs and K-type stars (bottom) from the G and F-type stars.  The systems below the black dashed line contain an M-dwarf star.  Systems marked by an asterisk contain a giant planet exterior to the snow line with a smaller planet interior to the snow line.}
    \label{fig:giant_systems}
\end{figure*}

 We now consider the architectures of the observed multi-planet systems around M,K,G and F-type stars.  Figure \ref{fig:giant_systems} shows all the multi-planet systems with mass measurements, that contain a giant planet.  We mark the location of the snow line from equation~(\ref{eq:Sun_snow}) with a blue vertical line and shade the $1\sigma$ bounds of the snow line location in light blue.  Stellar mass increases from bottom to top.  The red dashed line separates the M-dwarfs and K-type stars (stars with a mass $\leq 0.8 \, M_{\star}$) from the G and F-type stars.  The systems below the black dashed line contain an M-dwarf star.


We find no systems around low mass stars that contain a giant planet exterior to the snow line with a smaller planet interior to the snow line.  We find five such systems around G and F-type stars. We mark these five systems with an asterisk in Figure \ref{fig:giant_systems}.  Three of the five M-dwarf systems were observed using microlensing techniques, which would not be sensitive to terrestrial planets in the habitable zone due to the small star-planet separation \citep{Park2006}.  Follow up observations will help determine if these systems are truly devoid of an inner terrestrial system.


Multi-planet systems that contain a giant planet are also less common around low-mass stars than they are around G-type stars.  Of the G-type systems that contain a giant planet, $~13\%$ are multi-planet systems whereas only $~8\%$ of M-dwarf systems with a giant planet contain planet companions.

\section{Discussion} \label{sec:Discussion}
While our argument assumes that giant planets are needed to form stable asteroid belts and that giant planets typically form exterior to the snow line, alternative formation theories for both asteroid belts and giant planets are worth mentioning.

Alternative theories for producing a stable asteroid belt exist, including a late stage population of the belt region of planetesimals that are scattered from other locations during terrestrial planet formation, which may not require a giant planet \citep{Raymond2017, Raymond2022}.   However, even if asteroid belts are able to form around M-dwarfs in the absence of exterior giant planets, the system will still require a mechanism to deliver the asteroids to the terrestrial planets.

\cite{Mercer2020} have proposed that giant planets around M-dwarfs form through disk fragmentation.  While this scenario may help explain giant planets found on longer orbits, it requires a disk with an initial mass of at minimum, $0.3 \, M_{\star}$.  Such massive disks around M-dwarfs have yet to be observed.


The lack of hot and warm Jupiters around M-dwarfs shown in Figure \ref{fig:giant_locations} may be because the gas disks around low-mass stars are not massive enough for efficient migration.  The migration timescale for a planet is inversely proportional to the surface density of the gas disk \citep{Tanaka_2002} and so, a low surface density profile correlates to a long migration timescale.  This may result in a planet failing to cross the snow line prior to the dissipation of the gas disk if it forms farther out.  Furthermore, efficient photoevaporation from highly active M-dwarfs can quickly deplete the inner disk regions which can completely cease migration \citep{Monsch2021}.  


\section{Conclusions} \label{sec:Conclusions}

We have shown that \textit{if} the emergence of life indeed requires asteroid impacts onto habitable planets, then exoplanets around M-dwarfs may encounter an additional challenge.  Giant planets exterior to the snow line radius are necessary for preventing planet formation there, creating an asteroid belt, and the delivery of the volatile rich asteroids to the inner terrestrial system after gas disk dispersal.  None of the currently observed planets in the habitable zone around M-dwarfs have a giant planet outside of the snow line radius and therefore are unlikely to have a stable asteroid belt.


By analyzing data from the Exoplanet Archive, we found that there are observed giant planets outside of the snow line radius around M-dwarfs, and in fact the distribution peaks there. This, combined with observations of warm dusts belts, suggests that asteroid belt formation may still be possible around M-dwarfs. However, we found that in addition to a lower occurrence rate of giant planets around M-dwarf stars, multi-planet systems which contain a giant planet are also less common around M-dwarfs than around G-type stars.
Lastly, we found a lack of hot and warm Jupiters around M-dwarfs, relative to the K, G and F-type stars, potentially indicating that giant planet formation and/or evolution does take separate pathways around M-dwarfs. 

We suggest that the typical planet formation mechanisms around low-mass stars may be unlikely to result in planetary architectures with a giant planet exterior to the snow line, an asteroid belt, and an inner terrestrial system in the habitable zone.  Of course our conclusion is not intended to discourage the search for life on exoplanets orbiting M-dwarfs. On the contrary, an unambiguous detection of biosignatures on such planets can serve to demonstrate that asteroid impacts may not be a necessary condition for the origin of life.

\begin{acknowledgements}
 We acknowledge support from NASA through grant 80NSSC21K0395.
\end{acknowledgements}

\bibliography{ref}{}
\bibliographystyle{aasjournal}

\end{document}